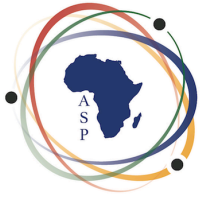 **African School of Fundamental Physics and Applications**

## Summary Report on the ASP2024 Learners Program

## April 15-19, 2024

Kétévi A. Assamagan[1], Abdelkarim Boskri[2], Kenneth Cecire[3], Mohamed Chabab[2], Christine Darve[4], Farida Fassi[5], Mounia Laassiri[1], Sanae Samsam[6], Janna Vischer[7]

*[1]Brookhaven National Laboratory USA, [2]Cadi Ayyad University Marrakech Morocco, [3]University of Notre dame USA, [4]European Spallation Source Sweden, [5]Mohammed V University Morocco, [6]INFN-Milan Italy, [7]*Friedrich-Alexander-Universität Erlangen-Nürnberg, G*ermany*

### 1) Introduction

The African School of Physics (ASP) [1] learners' program is designed to motivate high school pupils, aka learners, to develop and maintain interest in physics and related applications. It was integrated into the ASP program in 2016 when ASP was held in Rwanda [2] and has been a part of the ASP activities since. It gained popularity and grew through ASP20218 [3] and ASP2022 [4]. The learners' program is carried out over a week, in parallel to the university students' activities during term ASP, and involves high schools located in the vicinity of the ASP venue. Up to forty high schools can participate in the program, five of which serve as venues, different on each day. A maximum of two thousand learners can participate in the program, drawn from the forty high schools considered. The selected learners from a subset of high schools congregate at different designated venues on different days, for a targeted four-hour physics engagements in the mornings and afternoons. The participating high schools and learners are selected by the ASP Local Organizing Committee (LOC), working with the regional and national authorities responsible for learners' education.  The scientific program, proposed by the International; Organizing Committee (IOC) and the LOC, in collaboration with the local authorities, consists of a set of hands-on activities to illustrate physics concepts through

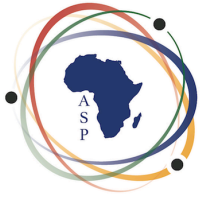

# African School of Fundamental Physics and Applications

engaging and challenging activities. Morocco was selected to host the 2024 edition of the African School of Physics in Marrakech in July 2024, as shown Figure 1.

Figure 1: Poster of the 8th edition of the African School of Physics, ASP2024.

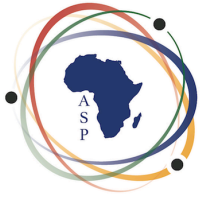 **African School of Fundamental Physics and Applications**

On April 15-19, 2024,  ASP2024 organized a program for learners from selected high schools in the vicinity of the venue. A second program for learners is planned for July 2024 during the main engagements of ASP2024. In that month, learners are normally in the major school break, and most of them are unavailable. To reach a larger number of learners, ASP2024 organizers decided to carry out the learners' program in April 2024 while learners are still accessible at the schools. Nevertheless, for the learners that will be available during the school break in July, a second learners' program will be arranged in parallel with the main ASP2024 activities.

In this paper, we report on the learners' program of April 15-19,2024, in the context of the 8[th] edition of the African School of Physics, ASP2024. Five high schools were selected to serve as venues on different days as shown in Table 1, along with the numbers of participants, high schools and selected learners, and the locations of the venues.

### 2) Scientific program

Up to four activity strands were planned to be performed in parallel at each venue: accelerator physics, astrophysics, particle physics, and detectors & instrumentation [5]. The learners were organized into groups assigned to different activities, and in some cases, rotations were implemented so that learners were exposed to the different sets of activities presented. Many of these activities required educational materials, most of which were shipped from abroad or broad in by the lecturers.

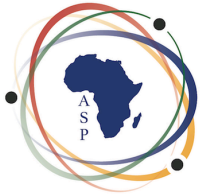

# African School of Fundamental Physics and Applications

| Day | Venue | Number of participating high schools | Number of learners & teachers | Distance from Marrakesh (km) |
|---|---|---|---|---|
| April 15 | Marrakesh | Lycée technique Mohammed VI | 238 learners, 31 teachers from 31 high schools | 0 |
| April 16 | Chichaoua | Lycée technique | 200 | 76 |
| April 17 | Al Haouz | Lycée Abtih | 200 | 50 |
| April 18 | El Kelaa-Sraghna | Lycée Tassout | 200 | 81 |
| April 19 | Lyçée Abdellah Ibrahim, Benguérir | Lycée Abdellah Ibrahim | 200 | 87 |

Table 1: Arrangement of the learners' program in April 2024

### 3) Scientific program

Up to four activity strands were planned to be performed in parallel at each venue: accelerator physics, astrophysics, particle physics, and detectors & instrumentation [5]. The learners were organized into groups assigned to different activities, and in some cases, rotations were implemented so that learners were exposed to the different sets of activities presented. Many of these activities required educational materials, most of which were shipped from abroad or broad in by the lecturers.

### a) Accelerator physics (to be done by S. Samsam and C. Darve)

We delivered two comprehensive lectures on accelerator physics and applications, complemented by engaging practical sessions.

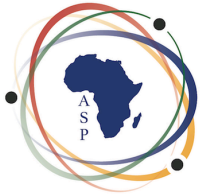

# African School of Fundamental Physics and Applications

After a short introduction of both Sanae and Christine, Sanae started her presentation in French and arabic. Her lecture has delved into the foundational aspects of accelerator physics, starting with accessible examples like cathode ray tubes to illustrate basic principles. We elucidated the five essential components of a particle accelerator: particles, energy, control, collision, and detection. Our discussion encompassed the fundamental roles of electric and magnetic fields in particle manipulation and control, elucidating the transition from theoretical understanding to practical implementation.

In order to ease the understanding, short animations and movies were used, both in French and in english. We have also referred to materials from the CEA platform, showing in french the use of particle accelerators.

To foster active participation, we employed an interactive approach, distributing four cards adorned with distinct symbols for a series of engaging Q&A sessions (See Fig.2). This method encouraged students to actively select the correct answers, promoting deeper comprehension and retention.

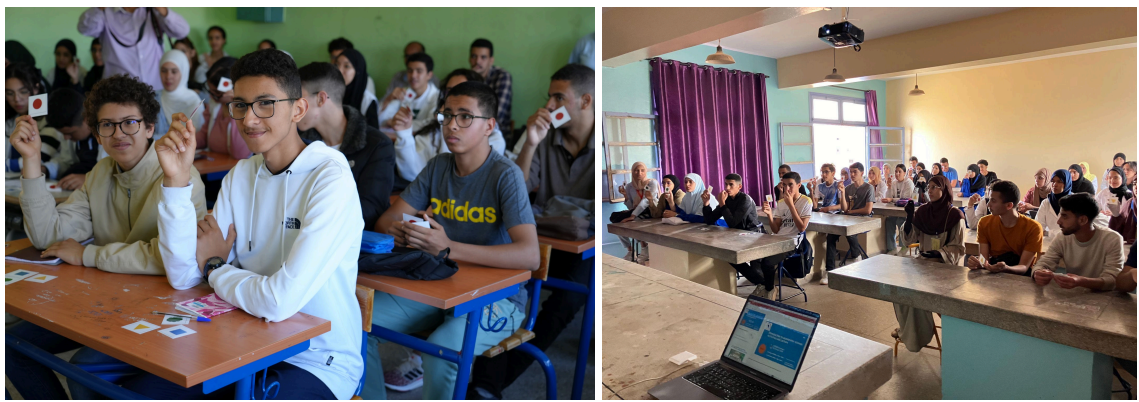

Figure 2: Interactive Q&A Session Sparks Enthusiasm in Accelerator Physics Course

Following the lectures, participants immersed themselves in hands-on activities. The first session involved constructing a Gauss cannon accelerator (See Fig. 3), a captivating exercise in converting magnetic energy into kinetic energy to propel a steel ball. This

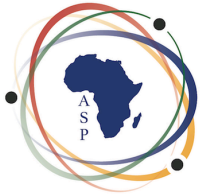

# African School of Fundamental Physics and Applications

experiment not only demonstrated energy conversion principles but also instilled a sense of inquiry as students explored concepts of magnetism through force and field measurements. The importance of simulation and programming emerged palpably as students recognized the necessity of thorough preparation before embarking on real-world accelerator projects.

In the subsequent session, students were tasked with assembling a linear and circular accelerator from scratch, offering a second hand experience of accelerator construction. Selected participants elucidated the five critical ingredients of an accelerator, from particle generation to detection, fostering a deeper appreciation for the role of electric and magnetic fields in particle manipulation. Through this hands-on endeavor, students not only had fun but also grasped the practical applications of accelerator technology, paving the way for future exploration and innovation. These hands-on activities allow learners to engage with the various concepts of linear particle accelerators (Fig. 3). One such interactive activity is the Gauss Cannon, designed to further engage participants. We provided a set of linear and magnetic components, which we assembled into a mock-up of a linear particle accelerator. Through this activity, we tested their understanding of energy transfer phenomena, forces, and acceleration concepts.

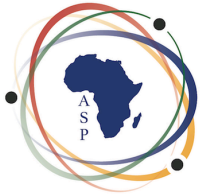

# African School of Fundamental Physics and Applications

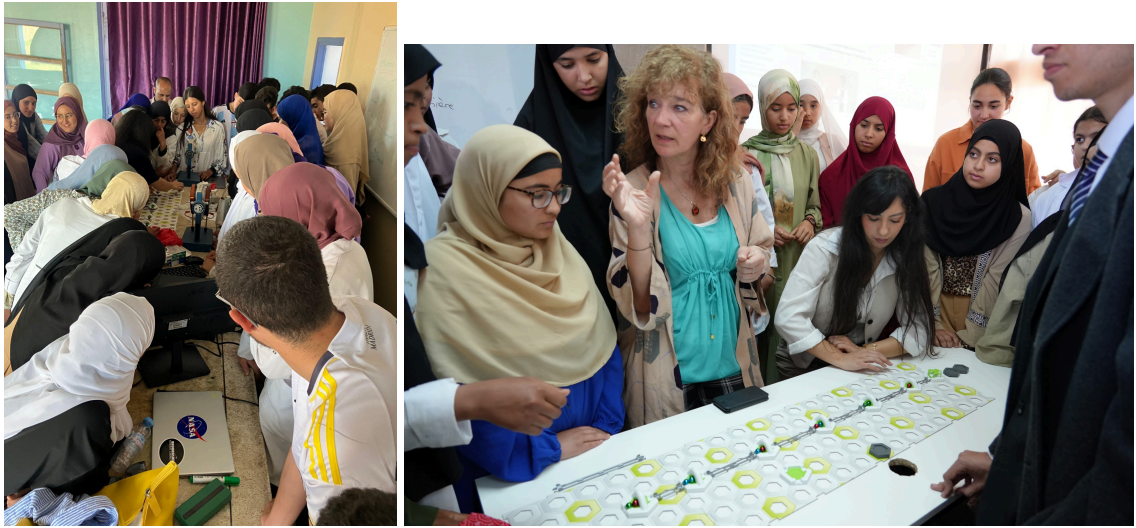

Figure 3: Gauss Cannon activities

After an engaging French introduction to particle accelerators, presented with French slides, Christine transitioned to English slides to review the fundamental laws of physics. She then introduced a newly launched Massive Open Online Course (MOOC), sponsored by the European Commission. This innovative course provides teachers with learning scenarios, short movies, and numerous animations, similar to the ones used in the introduction. Titled "Accelerate your Teaching," the MOOC is accessible for free on the EU SchoolNet platform. Its content covers the foundations of particles, magnetism, forces, light, and electromagnetism.

During the presentation, numerous URLs and links were shared to allow motivated students to delve deeper into the knowledge transferred from large research infrastructures. Additionally, APS and CERN brochures were distributed, and shown digitally, along with references to its 70th anniversary, inviting participants to join Open Source Zoom meetings.

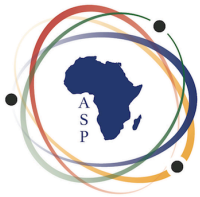

# African School of Fundamental Physics and Applications

Two main case studies were presented to illustrate the phenomenology behind particle accelerators. The first focused on Colliding Physics, opening learners to an understanding of the laws of the universe. The second case study explored Scattering Communities using Photons and Neutrons Sources. These examples aimed to demonstrate the practical applications of physics and the importance of scientific inquiry. They encouraged participants to question the world around them and highlighted the role of large-scale scientific infrastructures in driving innovation through international collaboration.

The presentation also emphasized the importance of English language proficiency, as it is the lingua franca of scientific communication.

To enhance the learning experience, hands-on activities were incorporated, allowing participants to explore the function of accelerating cavities and steering magnets through in-person exercises (Fig. 4). The concept of acceleration was introduced through role-playing, with learners embodying accelerating cavities and magnets in a control room, providing energy to simulated electrons or protons.

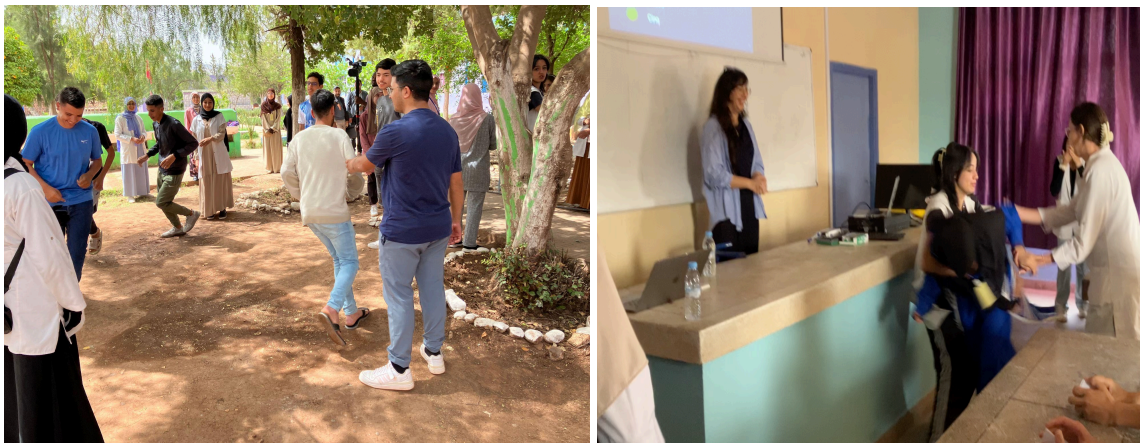

Figure 4: Building a human synchrotron and having fun in the classroom accelerating electrons

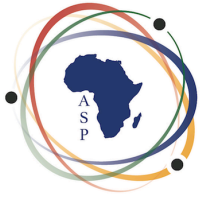

# African School of Fundamental Physics and Applications

⇒ The presentations are available to the public: "Introduction to accelerator science: Activities and Lecture", Apr. 15-17, '24 and **Part 1 and  Part 2**

### b)  Astrophysics

The astrophysics activity consisted in a 30-minute presentation—for a maximum of 25 learners accommodated in a makeshift planetarium. The activity was repeated several times to different groups of learners, in the mornings and afternoons of April 15 & 17. During the planetarium show, the instructors explained the movements of celestial bodies, astronomical observations, and the interpretation of the sky during different seasons; an example is shown in Figure 5.

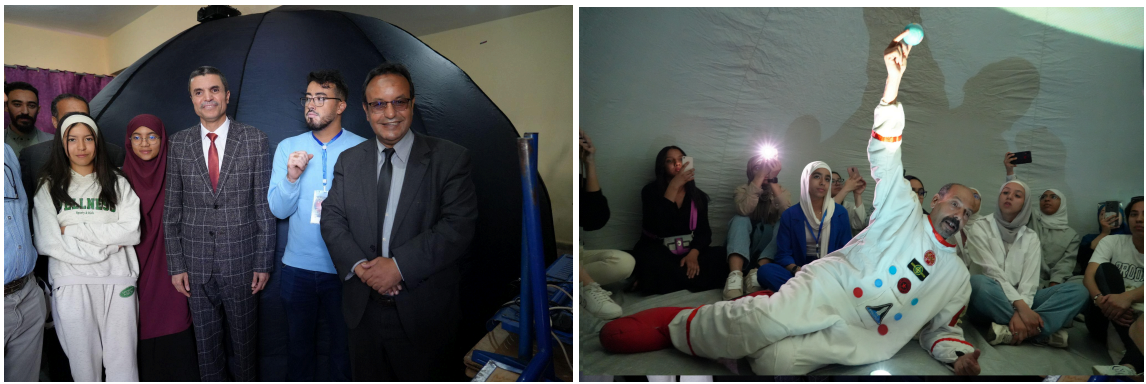

Figure 5: Regional education authorities, astrophysics instructor and learners in front of a makeshift planetarium (left). An instructor and learners inside the planetarium (right).

### c)  Particle physics

Our workshop aims to introduce the learners to the standard model of particle physics and to familiarize them with the scientific methods of measuring the tiniest particles.

As an introduction we introduce the learners to the building blocks of matter, zooming in from molecules to atoms and their smallest constituents: electrons and quarks. These

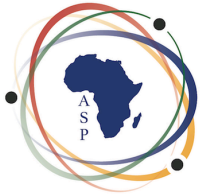

# African School of Fundamental Physics and Applications

topics might already be partially known, depending on the age group and background of the learners. We introduce them to the concept of elementary, undividable particles.

In our first hands-on activity, the learners form small groups of 3-6 students and are asked to sort a set of cards[1], where each card represents one elementary particle that has been discovered by physicists, as shown in Figure 6. They are supposed to sort the particle cards by their properties or into groups, similar to e.g the periodic table. For example, learners may initially order the cards by lifetime or mass and then according to electric or weak charge. We explain that we are looking for their organization and what they might learn from it, so whatever ordering they come up with is, by definition, valid.

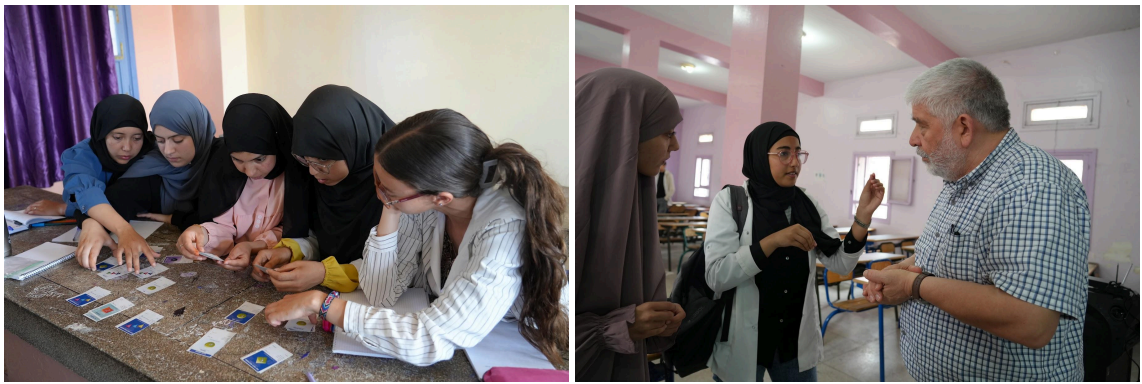

Figure 6: A group of students working together to order the elementary particles (left). Students asking individual follow-up questions after the workshop (right).

We discussed the different elementary particles and their properties, as well as the four fundamental forces and their impact on our daily lives.

---

[1] https://quarknet.org/data-portfolio/activity/shuffling-particle-deck

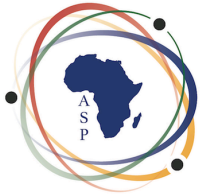

# African School of Fundamental Physics and Applications

To illustrate how physicists are able to measure particles that are too small to visually observe, the determination of the size of marbles[2] was used: By randomly colliding marbles of a certain size with other marbles and counting the number of hits, the learners were able to create a histogram of numbers of hits out of ten rolls and from there derive the size of the marbles based on statistics and a simple probability proposition. As in large physics projects, a large number of collisions is necessary to obtain sufficient statistics for a physical conclusion, which is achieved by the collective effort of the group.

We introduced the learners briefly to the idea of particle accelerators and the concept of CERN and the LHC. To visualize the collision and creation of new particles the learners pretended to be particles themselves and reenacted a particle collision, with individual learners taking on the roles of protons, a Z boson, and muons in the LHC.

In the final part of the workshop, the learners were able to measure real elementary particles, i.e. muons, created in the upper atmosphere by primary cosmic rays using our small, portable Cosmic Ray Muon Detector[3]. After a brief introduction of the setup and detector components the learners were able to participate in muon measurements. The most common measurement was of the s from where  the cosmic ray muons are coming by counting events in vertical and horizontal directions. Other measurements were also performed as time permitted.

We concluded our workshop with a Q&A session in which the learners were able to ask questions about particle physics, physics in general and life as a physicist.

---

[2] https://quarknet.org/data-portfolio/activity/rolling-rutherford
[3] https://quarknet.org/sites/default/files/cf_6000crmdusermanual-small.pdf

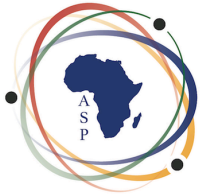

# African School of Fundamental Physics and Applications

The unifying themes of the workshop were "What do we know?" and "How do we know?", leaving room to discuss what we do not know or are not sure about. In the first part, learners are engaged by an introduction to the elementary particles and, to a limited extent, how different they are from our macroscopic experience of matter and forces, opening a pathway to learning about quantum physics. In the second part, learners see how what we have learned about the particle world is based on experimentation, data, and analyses that enable physicists to indirectly measure states that are far too small to measure directly. Hopefully, in the end, learners can appreciate that the characteristics of this unseen world are "strange but true."

### d) Detectors and Instrumentation

We offered four hands-on lab activities to the learners:

- Geiger Counter – To correlate decay events (from over-the counter thoriated lantern mantles and TIG welding rods) as a function of distance from an inexpensive detector.
- Spectrophotometry – To catalog substances by their reflectance spectra and to use this database to identify unlabeled materials.
- Ohm's Law – Empirically determine the relationship between resistance, current, and voltage and to predict the resistance of an unlabeled resistor.
- Center of Mass – Mathematically determine the balancing point of 14 and 15-block Jenga cantilevers and physically build those piles.

The equipment needed for the Geiger counter, spectrophotometry and Ohm's law were shipped from the U.S. Unfortunately, they did not arrive in Morocco in time. Only the center-of-mass exercise was performed. First, a short introduction about the concept of center-of-mass was presented, supported by the mathematical formalism for computing the coordinates of center-of-mass in a reference frame. Learners were involved in the

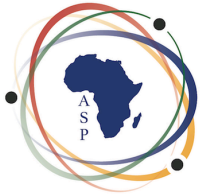

# African School of Fundamental Physics and Applications

derivations. Then, we proceeded to practical applications in building cantilevers using Jenga blocks. Here, we discussed the maximum size of the cantilever and the condition for stability. Finally, we discussed practical applications of cantilevers in buildings, bridges, etc. Figure 7 shows learners successfully building 15-block Jenga cantilevers, with a success rate better than 20%.

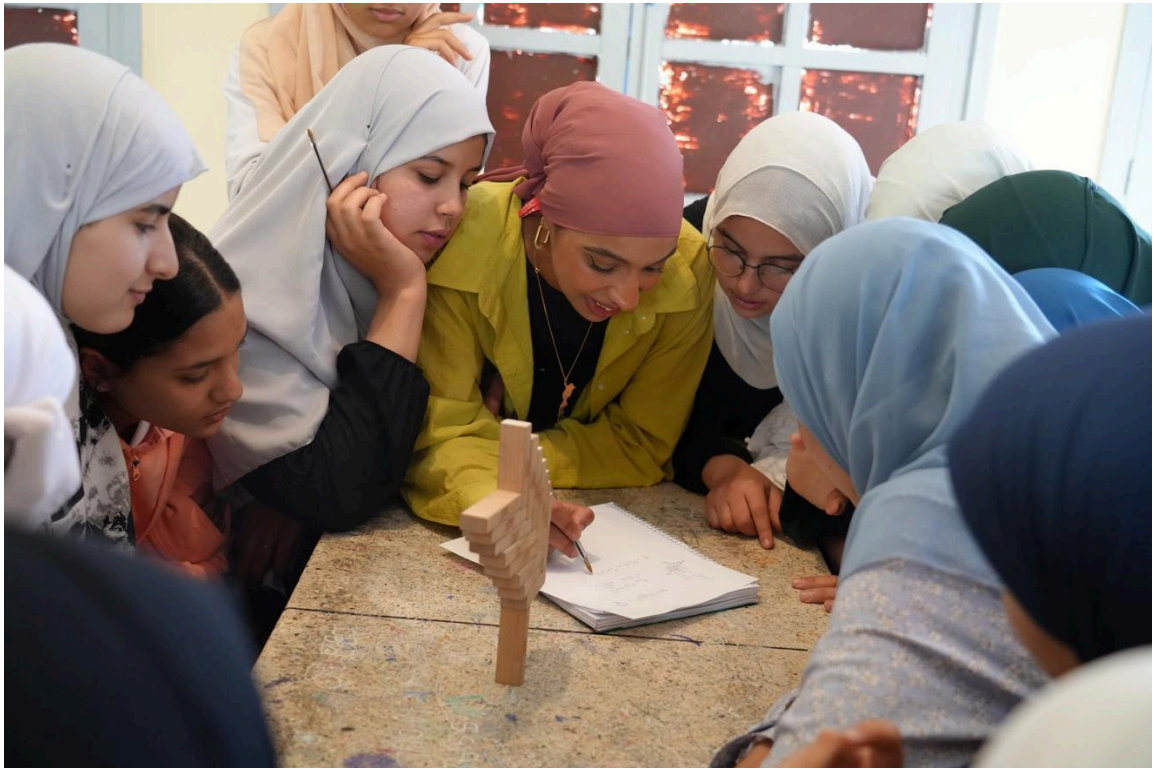

Figure 7: Lecturer, Dr. Mounia Laassiri (center), discussing the mathematical formalism of the concept of center-of-mass and the condition of stability of the 15-block Jenga cantilever that the learners succeeded in building.

A short video of the event is available in Ref. [6].

4) **Feedback, outlook, and conclusions**

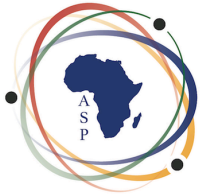 # African School of Fundamental Physics and Applications

The ASP2024 program for learners was organized as a team effort between the ASP IOC, the ASP2024 LOC, ASP lecturers and regional educational authorities in Marrakesh. Five high schools were selected as venues, one for each day during the period of April 15-19, 2024. The date of the event was chosen to maximize the learners' attendance; the event drew the participation of about 1000 learners selected from different high schools and organized to congregate at a designated venue. Up to four sets of physics programs, in accelerators, astrophysics, particle physics, and detector & instrumentation, were carried out parallel at the venue. Learners had opportunities to participate in various hands-on activities designed to motivate and sustain their interest in physics and related applications. Feedback received from the local authorities, teachers, and learners themselves suggest that the event was a successful one, with a great interest to be organized in the future. The main engagements of ASP2024 are planned for July 2024 where the scientific program will include activities for university students and high school teachers. Although learners will be on their major school break in July, we plan to organize the learners' program again, in parallel with the other activities, for learners that will be available. Most of the education materials used in these learners' programs will be donated to the participating high schools for future use.

**Acknowledgements**

We would like to thank the ASP sponsors for their continued support without which such events cannot be performed. We also thank the regional education authorities in Marrakesh, particularly Messrs. Belhind, Elmnaoui et  Belhou, for supporting the events and contributing to its success. We also thank the chairs of the LOC coordinating the engagements between the IOC, the lecturers, the authorities, and the high schools. Special thanks to Gilbert Tékouté for photographing, filming, and recording the event.

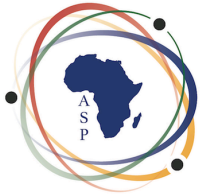 **African School of Fundamental Physics and Applications**

We would like to thank Dr. Abdelatif Zridi for the planetarium show on April 15, and Dr. Abd Errahmane Ben Jaddi and Dr Marwane Leili of the Sahara Astronomy Association for the planetarium show on April 17, 2024.

Kétévi A. Assamagan and Mounia Laassiri would like to acknowledge and thank David Biersack, Johanthan Ullman, and Daniel Trieu for putting together the four activities in detectors & instrumentation and offering tutorials to carry them successfully with the learners.